\documentclass[pdftex,twocolumn,epj]{svjour}
\usepackage[T1]{fontenc}
\usepackage[utf8]{inputenc}
\usepackage{bm}
\usepackage{latexsym}
\usepackage{graphics}
\usepackage{times}
\usepackage{palatino}
\usepackage{epsf}
\usepackage{amsmath}
\usepackage{amssymb}
\usepackage{latexsym}
\usepackage{calc}
\usepackage{color}
\usepackage{shadow}
\usepackage{epsfig}
\usepackage[usenames,dvipsnames]{xcolor}
\usepackage[toc,title,page]{appendix}

\def\bea{\begin{eqnarray}}
\def\eea{\end{eqnarray}}
\def\ben{\begin{equation}}
\def\een{\end{equation}}
\def\benu{\begin{enumerate}}
\def\enu{\end{enumerate}}

\def\dulr{{\underline{\underline{\bf r}}}}
\def\dulq{{\underline{\underline{\bf q}}}}

\def\1var{(\bx_1...\bx\N)}

\def\br{{\bf r}}

\def\bx{{\br t}}

\def\dulr{{\underline{\underline{\bf r}}}}

\def\bea{\begin{eqnarray}}
\def\eea{\end{eqnarray}}
\def\ben{\begin{equation}}
\def\een{\end{equation}}
\def\benu{\begin{enumerate}}
\def\enu{\end{enumerate}}

\def\dulr{{\underline{\underline{\bf r}}}}

\begin{document}
\title{Shedding Light on Correlated Electron-Photon States using the Exact Factorization}
\author{Ali Abedi\inst{1,2}\thanks{aliabedik@gmail.com\vspace{2pt}} \and Elham Khosravi\inst{1,2}\thanks{elham.etn@gmail.com\vspace{2pt}} \and Ilya V. Tokatly\inst{1,2,3}\thanks{ilya.tokatly@ehu.es}}                    
\institute{Nano-Bio Spectroscopy group, Departamento de F\'isica de
  Materiales, Universidad del Pa\'is Vasco UPV/EHU,\\ Av.~Tolosa 72, E-20018 San
  Sebasti\'an, Spain\label{1} 
  \and Donostia International Physics Center (DIPC), Manuel de Lardizabal 5, E-20018 San Sebasti\'an, Spain\label{2} 
  \and IKERBASQUE, Basque Foundation for Science, E-48011 Bilbao, Spain\label{3}}

\date{Received: date / Revised version: date}
%
\abstract{The Exact Factorization framework is extended and utilized to introduce the electronic-states of correlated electron-photon systems. The formal definitions of an exact scalar potential and an exact 
vector potential that account for the electron-photon correlation are given. Inclusion of these potentials to the Hamiltonian of the uncoupled electronic system leads to a purely electronic Schr\"odinger equation 
that uniquely determines the electronic states of the complete electron-photon system. For a one-dimensional asymmetric double-well potential coupled to a single photon mode with resonance frequency, we investigate 
the features of the exact scalar potential. In particular, we discuss the significance of the step-and-peak structure of the exact scalar potential in describing the phenomena of photon-assisted delocalization and 
polaritonic squeezing of the electronic excited-states. In addition, we develop an analytical approximation for the scalar potential and demonstrate how the step-and-peak features of the exact scalar potential are captured 
by the proposed analytical expression. 
\PACS{
      {PACS-key}{discribing text of that key}   \and
      {PACS-key}{discribing text of that key}
     } 
} 
\maketitle

\section{Introduction}
Rapid progress in the fields of cavity and circuit quantum electrodynamics (QED) has given us a possibility to study how many-electron systems interact with quantum light. The interplay between photons
and electrons plays a key role in many fascinating processes in atoms inside optical cavities in cavity-QED~\cite{RaiBruHar2001,MabDoh2002,Walter2006}, or mesoscopic systems such as superconducting qubits and quantum dots embedded in transmission line resonators in circuit-QED \cite{Wallraff2004,Blais2004,Frey2012,Delbecq2011,Petersson2012,Liu2014}. 
Furthermore, in the case of molecules a strong coupling of molecular states to microcavity photons has been achieved experimentally \cite{Schwartz2011,Ebbesen2016}, and the modification of photo-chemical landscapes, the charge and energy transport by cavity vacuum fields have been reported \cite{Hutchison2012,Orgiu2015,Zhong2017}. The progress in experiments triggered theoretical activities that address ``chemistry-in-cavity'' problem. Indeed, the corresponding processes cannot be captured properly within the usual classical approximation for the light as the system now includes new quantum degrees of freedom of photons and the concept of electron-photon correlation comes in as a new player influencing the electronic states of the system. In the last few years, 
several theoretical approaches have been put forward to describe molecular systems in quantum cavities. These include both mapping to simplified few-level quantum optics models \cite{KowBenMuk2016,GalGarFei2015,GalGarFei2016,GalGarFei2017}, and the 
cavity-QED generalizations of {\it ab initio} electronic structure methods, such as (TD)DFT~\cite{Tokatly2013PRL,Ruggenthaler2014PRA,Pellegrini2015PRL,R15,Flick2015,Flick2017a,FSR18,DFRR17}, or the Hedin equations framework in the Green functions theory \cite{TreMil2015}.

As long as the light can be treated classically,
the electronic many-body states are fully described by the non-relativistic   electronic Schr\"odinger Equation (SE), i. e., 
\ben
\label{eq:ese}
 \hat{H}_{e} \phi^j(\dulr)= E_j\phi^j(\dulr),
\een
with the Hamiltonian
\begin{equation}
\label{eq:e_ucH}
\hat{H}_{e}=\hat{T}_e+\hat{V}+\hat{W}_{ee}
\end{equation}
where $\hat{T}_e$, $\hat{V}$ and $\hat{W}_{ee}$ are the usual kinetic energy, the external potential, and the Coulomb interaction energies of the electrons, respectively. The Hamiltonian $\hat{H}_{e}$ acts 
on the electronic coordinates collectively denoted by ${\dulr} \equiv {r_1, r_2, \ldots, r_{N_e}}$. At this level of theory only the electrons are treated quantum mechanically and only the coordinates of electrons 
appear as arguments of the wavefunction. However, if the electronic system is embedded in a microcavity the presence of quantum electromagnetic degrees of freedom (photons) can modify the electronic states 
significantly. The complete description of the quantum states of matter that is now coupled to the photons, 
in principle, can be provided by the SE of the multi-component system of electrons and photons, 
\ben
\label{eq:se}
 \hat{H}_\text{tot} \Psi^j(\dulr,\dulq)= E_j\Psi^j(\dulr,\dulq),
\een
with the Hamiltonian
\begin{equation}
\label{H_tot}
\hat{H}_\text{tot}=\hat{H}_{e}+\hat{H}_{\text{EM}},
\end{equation}
which now includes an additional term, $\hat{H}_{\text{EM}}$, that is resulted from the quantization of the electromagnetic field to properly account for quantum features of the radiation field and the electron-photon correlation.
A detailed derivation of $\hat{H}_{\text{EM}}$ will be presented shortly in the following section. Here, the degrees of freedom of $N_p$ cavity modes are collectively represented 
by ${\dulq} \equiv {q_1, q_2, \ldots, q_{N_p}}$. In the length gauge, the ``electromagnetic coordinate'' $q_{\alpha}$ corresponds to the amplitude of electric displacement in the $\alpha$-mode of the cavity (see section~\ref{section:QED_Ham} for more details).

A numerically exact solution of the complete electron-photon SE, Eq.~(\ref{eq:se}), can only be obtained for small systems, hence, an accurate description of the electronic 
states of matter in the presence of photons requires an efficient treatment of electronic many-body problem while accounting for the electron-photon correlation. Here, an important question is whether or not the information 
on the electronic states that are coupled to photons can be obtained from pure electronic states ${\Phi^j(\dulr)}$ rather than the more complicated electron-photon states ${\Psi^j(\dulr,\dulq)}$. And if yes, what Hamiltonian 
gives such electronic states? How does it differ from the uncoupled electronic Hamiltonian of Eq.~(\ref{eq:e_ucH})? 

In this work, we will address these questions and will demonstrate that the answer to the first question is indeed yes. By utilizing the Exact Factorization (EF) framework \cite{hunter_IJQC1975,GG_PTRSA2014,AMG_PRL2010}
we will show how pure electronic states, ${\Phi^j(\dulr)}$, can provide us with important information such as the exact electronic many-body densities and current densities equivalent to those obtained from the complete 
electron-photon states ${\Psi^j(\dulr,\dulq)}$. We furthermore, present the additional purely electronic potentials that are needed to be included in the electronic Hamiltonian in order to account for the electron-photon correlation in a formally exact way. In addition, we derive analytical expressions of these potentials for electronic states of a single electron system in an asymmetric double-well potential that is coupled to 
a single-photon mode of a cavity with a resonance frequency. Furthermore, we will show how well our analytical expressions match the potential obtained from the numerical solution of the SE. We  demonstrate that in the resonance regime the effective electronic potential for excited states demonstrate clear peak and step structures, which are responsible, respectively, for the polaritonic squeezing  of the intra-well states and the photon-assisted inter-well tunneling.
 
Here, we only study the stationary states. However our results have a direct relevance for understanding the dynamics of electron-photon systems and thus for the development of QED-TDDFT \cite{Tokatly2013PRL,Ruggenthaler2014PRA,Flick2015}. In fact, the step-and-peak structure of the electronic potential for the stationary excited states should show up in the effective time-dependent potential to account for the charge transfer processes supplemented with the photon emission/absorption.

\section{Quantization of electromagnetic field: Hamiltonian for cavity QED}
\label{section:QED_Ham}
The main object of the cavity/circuit QED is a system of non-relativistic
electrons interacting with electromagnetic modes of a microcavity. The QED regime assumes that both electrons and the electromagnetic
field are treated quantum mechanically. However, 
to  understand better the structure of the quantum theory 
it is instructive to analyze first the classical dynamics of the system. 

Our starting point is the Maxwell equations for the transverse part
of the electromagnetic filed
\begin{eqnarray}
& &\nabla\times{\bf E}_{\perp}  =  -\frac{1}{c}\partial_{t}{\bf B},\label{Maxwell-EB-1}\\
& &\nabla\times{\bf B} \, \, \, =  \frac{1}{c}\partial_{t}{\bf E}_{\perp}+\frac{4\pi}{c}{\bf j}_{\perp},\label{Maxwell-EB-2}
\end{eqnarray}
where ${\bf E}_{\perp}({\bf r},t)$ is the transverse electric field
with $\nabla\cdot{\bf E}_{\perp}=0,$ and ${\bf j}_{\perp}({\bf r},t)$
is the transverse part of electron current that enters as a source
of the radiation field. In general for $N_e$ electrons moving along
trajectories ${\bf r}_{j}(t)$ the current is defined as follows 
\begin{equation}
{\bf j}({\bf r},t)=e\sum_{j=1}^{N_e}\dot{{\bf r}}_{j}(t)\delta({\bf r}-{\bf r}_{j}(t)).\label{current}
\end{equation}
In a typical cavity QED setup, the motion of electrons is bounded to
a region around some point ${\bf r}_{0}$ inside the cavity, which
is much smaller than the cavity size and thus much smaller than the
characteristic wavelength $\lambda$ of the field. The condition $|{\bf r}_{j}(t)-{\bf r}_{0}|\ll\lambda$
justifies the replacement of ${\bf r}_{j}\mapsto{\bf r}_{0}$ in the
arguments of the $\delta$-functions
\begin{equation}
{\bf j}({\bf r},t)=e\sum_{j=1}^{N}\dot{{\bf r}}_{j}(t)\delta({\bf r}-{\bf r}_{0})=\partial_{t}{\bf P}({\bf r},t)\label{current-P}
\end{equation}
where we introduced the polarization vector ${\bf P}({\bf r},t)=e{\bf R}(t)\delta({\bf r}-{\bf r}_{0})$
with ${\bf R}=\sum_{j=1}^{N}{\bf r}_{j}$ being the center-of-mass
coordinate of the electrons. This corresponds to the dipole approximation
that is fulfilled with a very high accuracy in most of the practical situations.
The transverse current entering the Maxwell equations is determined
by the transverse projection of the polarization vector 
\begin{equation}
{\bf P}_{\perp}({\bf r},t)=e{\bf R}(t)\delta^{\perp}({\bf r}-{\bf r}_{0})=\frac{e}{4\pi}\nabla\times\left(\nabla\times\frac{{\bf R}(t)}{|{\bf r}-{\bf r}_{0}|}\right)\label{P-transverse},
\end{equation}
where we have used the identity $\delta^{\perp}({\bf r}-{\bf r}_{0})=\frac{1}{4\pi}\nabla\times\left(\nabla\times\frac{1}{|{\bf r}-{\bf r}_{0}|}\right)$.
In a quantum theory, the Maxwell equations (\ref{Maxwell-EB-1}), (\ref{Maxwell-EB-2})
should become Heisenberg equations for the corresponding field operators.
The quantum operator algebra can be revealed by representing the classical
theory in a Hamiltonian form. To do so, we introduce a new electric
variable -- the displacement vector 
\begin{equation}
{\bf D_{\perp}}={\bf E}_{\perp}+4\pi{\bf P}_{\perp},\label{D}
\end{equation}
and rewrite the Maxwell equations(\ref{Maxwell-EB-1}), (\ref{Maxwell-EB-2})
as follows 
\begin{eqnarray}
& &\partial_{t}{\bf B} \, \, \, \, =  -c\nabla\times({\bf D}_{\perp}-4\pi{\bf P}_{\perp}),\label{Maxwell-DB-1}\\
& &\partial_{t}{\bf D}_{\perp}  =  c\nabla\times{\bf B}.\label{Maxwell-DB-2}
\end{eqnarray}
These equations demonstrate a clear Hamiltonian structure. Indeed,
by considering the standard energy of the transverse electromagnetic
field
\begin{eqnarray}
H_{\text{EM}} & = &\frac{1}{8\pi}\int d{\bf r}\left[{\bf E}_{\perp}^{2}+{\bf B}^{2}\right] \nonumber\\
               & = &\frac{1}{8\pi}\int d{\bf r}\left[({\bf D}_{\perp}-4\pi{\bf P}_{\perp})^{2}+{\bf B}^{2}\right],\label{He-m}
\end{eqnarray}
and imposing the following commutation relations
\begin{equation}
[B^{i}({\bf r}),D_{\perp}^{j}({\bf r}')]=-i\,4\pi c \, \varepsilon^{ijk}\partial_{k}\delta({\bf r}-{\bf r}'),\label{BD-commut}
\end{equation}
we recover the Maxwell equations from the canonical Heisenberg equations
\begin{eqnarray}
& &\partial_{t}{\bf B} \, \, \, =  i[H_{\text{EM}},{\bf D}_{\perp}],\label{Maxwell-H-1}\\
& &\partial_{t}{\bf D}_{\perp}  =  i[H_{\text{EM}},{\bf B}].\label{Maxwell-H-2}
\end{eqnarray}

An important outcome of this analysis is that the proper conjugated
Hamiltonian variables for the electromagnetic field are the magnetic
field ${\bf B}$ and the electric displacement ${\bf D}$. 

Let us introduce cavity modes as a set of normalized transverse eigenfunctions
${\bf E}_{\alpha}({\bf r})$ of the wave equation inside a metallic
cavity $\Omega$
\begin{eqnarray*}
& &c^{2}\nabla^{2}{\bf E}_{\alpha}({\bf r}) \, \, =  \omega_{\alpha}^{2}{\bf E}_{\alpha}({\bf r}),\quad{\bf r}\in\Omega\\
& &({\bf n\times{\bf E}_{\alpha}})\vert_{\partial\Omega}  =  0,
\end{eqnarray*}
where ${\bf n}$ is a unit vector normal to the cavity surface $\partial\Omega$.
Now all transverse functions in the Hamiltonian (\ref{He-m}) can
be expanded in the cavity modes
\begin{eqnarray}
{\bf D}_{\perp}({\bf r}) & = & \sum_{\alpha}d_{\alpha}{\bf E}_{\alpha}({\bf r}),\label{D-expansion}\\
{\bf B}({\bf r}) & = & \sum_{\alpha}b_{\alpha}\frac{c}{\omega_{\alpha}}\nabla\times{\bf E}_{\alpha}({\bf r}),\label{B-expansion}\\
{\bf P}_{\perp}({\bf r}) & = & e\sum_{\alpha}\left({\bf E}_{\alpha}({\bf r}_{0})\cdot{\bf R}\right){\bf E}_{\alpha}({\bf r}).\label{P-expansion}
\end{eqnarray}
Here the expansion coefficients $d_{\alpha}$ and $b_{\alpha}$ are,
respectively, the quantum amplitudes of the electric displacement
and the magnetic field in the $\alpha$-mode. Note that Eq.~(\ref{B-expansion}) ensures that the magnetic field satisfies the proper boundary condition $({\bf B}\cdot {\bf n})\vert_{\partial\Omega}= 0$. By inserting the above expansions into Eqs.~(\ref{He-m}) and~(\ref{BD-commut}) we obtain the following Hamiltonian
\begin{equation}
H_{\text{EM}}=\frac{1}{8\pi}\sum_{\alpha}\left[\left(d_{\alpha}-4\pi e{\bf E}_{\alpha}({\bf r}_{0})\cdot{\bf R}\right)^{2}+b_{\alpha}^{2}\right],\label{H-bd}
\end{equation}
and the commutation relations for the field amplitudes
\begin{equation}
[b_{\alpha},d_{\beta}]=-i4\pi\omega_{\alpha}\delta_{\alpha\beta}.\label{bd-commut}
\end{equation}

Finally we rescale the electric displacement and the magnetic field
amplitudes
\begin{equation}
d_{\alpha}=\sqrt{4\pi}\omega_{\alpha}q_{\alpha},\quad b_{\alpha}=\sqrt{4\pi}p_{\alpha},\label{pq-def}
\end{equation}
so that the new variables $q_{\alpha}$ and $p_{\alpha}$ satisfy
the standard coordinate-momentum commutation relations $[p_{\alpha},q_{\beta}]=-i\delta_{\alpha\beta}$,
while the Hamiltonian (\ref{H-bd}) reduces to that of a set of shifted
harmonic oscillators
\begin{equation}
H_{\text{EM}}=\frac{1}{2}\sum_{\alpha}\left[p_{\alpha}^{2}+\omega_{\alpha}^2\left(q_{\alpha}-\frac{\bm{\lambda}_{\alpha} \cdot \bf R}{\omega_{\alpha}}\right)^{2}\right],\label{Hem-PZW}
\end{equation}
where the ``coupling constant'' $\bm{\lambda}_{\alpha}$ is related
to the electric field of the $\alpha$-mode at the location of the
electron system
\begin{equation}
\bm{\lambda}_{\alpha}=\sqrt{4\pi}e{\bf E}_{\alpha}({\bf r}_{0}).\label{lambda}
\end{equation}

Equation (\ref{Hem-PZW}) corresponds to the description of quantum
electromagnetic field and the electron-photon coupling in a so called
Power-Zienau-Woolley (PZW) gauge \cite{PowZie1959,Woolley1971,BabLou1983}. The total Hamiltonian for the combined system of electrons and the field is a sum of $H_{\text{EM}}$ and the standard Hamiltonian of a non-relativistic
many-electron system
\begin{equation}
\hat{H}_\text{tot}=\hat{H}_{e}+\hat{H}_{\text{EM}}\label{H-fin},
\end{equation}
where $\hat{H}_{e}$ is the electronic Hamiltonian of Eq.~(\ref{eq:e_ucH}) 
in the absence of the photon field. This Hamiltonian is commonly used
as the starting point in the first-principles approaches to the cavity
QED \cite{Tokatly2013PRL,Ruggenthaler2014PRA,Pellegrini2015PRL,Flick2015,Flick2017a}. 

\section{Exact factorization of the complete electron-photon wavefunction} 
The framework of the exact factorization (EF) for static \cite{GG_PTRSA2014,cederbaum_jcp2013,hunter_IJQC1975} and time-dependent problems \cite{AMG_PRL2010,AMG_JCP2012,AMG_JCP2013} was originally developed to go beyond 
the Born-Oppenheimer treatment of multicomponent systems of electrons and nuclei. Consequently, the original presentation of the framework provides an exact separation of the complete electron-nuclear wavefunction 
as a product of a marginal nuclear wavefunction and a conditional electronic wavefunction that parametrically depends on the nuclear configuration. As there is no approximation involved in developing this framework and 
the two subsystems are treated on the same footings, in principle, the EF can be extended to exactly factorize any multicomponent many-body wavefunction. In general, the choice of marginal and conditional wavefunctions 
is arbitrary and depends on the setting of the problem and its applications. Within the EF approach, the expressions of the coupling potentials 
that account for the exact correlation between the two subsystems are given explicitly and the conditional wavefunction satisfies a partial normalization condition. While the equation of motion (EoM) of the marginal wavefunction has an appealing form of a (TD)SE that includes an scalar and a 
vector potential, the EoM of the conditional wavefunction is non-linear and depends on both conditional and marginal wavefunctions.  
The EF approach has grown steadily over the past couple of years and has been implemented for fundamental investigations and method developments in various fields such as molecular dynamics~\cite{AASG_PRL2013,AASG_MP2013,AAG_EPL2014,AAG_JCP2014,AASMMG_JCP2015,SAG_JPCA2015,AMAG_JCTC2016,CA_JPCL2017,MATG_JPCL2017,SASGV_PRX2017,HLM_JPCL2018}, 
geometric phases~\cite{MAKG_PRL2014,RTG_PRA2016,RG_PRL2016} and strong-field dynamics~\cite{AMG_PRL2010,AMG_JCP2012,SAMYG_PRA2014,SAMG_PCCP2015,KAM_PRL2015,KARM_PCCP2017,SG_PRL2017}.

In this section, we present a generalization of the EF approach for the problem of correlated electron-photon states. As our derivation follows closely the procedure given in~\cite{GG_PTRSA2014} for the correlated electron-nuclear 
states, here we only present the final outcomes of the derivation and refer the readers to the reference~\cite{GG_PTRSA2014} for more details~\footnote{After submitting 
this manuscript we became aware of a recent unpublished work on the dynamical aspects of the light-matter interaction using the time-dependent EF frameworkin ~\cite{HARM18}.}.

Within the EF framework, the ($j$-th) correlated electron-photon state that is an {\it exact}  eigenstate of the complete electron-photon SE~(\ref{eq:se}), $\Psi^j(\dulr,\dulq)$, can be written as a single product, 
of an electronic wavefunction, $\Phi^j(\dulr)$, and a photonic wavefunction parameterized by the electronic coordinates, $\chi^j_{\dulr}(\dulq)$, i. e.,
\ben
\label{eq:EF1} 
\Psi^j(\dulr,\dulq)=\Phi^j(\dulr)\chi^j_{\dulr}(\dulq),
\een
that satisfies the partial normalization condition (PNC) 
\ben
\label{eq:PNC} 
\int d\dulq \, \vert\chi^j_{\dulr}(\dulq)\vert^2=1  ~\text {for every}~~\dulr.
\een
Here, we emphasize again on the vital role of the PNC in making this product physically meaningful. Indeed, it is possible to come up with a lot of different decompositions that satisfy Eq.~(\ref{eq:EF1}) but {\it do not} 
fulfill the PNC. As it was discussed previously, for instance in Ref.~\cite{AMG_JCP2013}, it is the PNC that makes the decomposition physically meaningful and unique up to a gauge-like transformation and allows for the interpretation of
a marginal probability amplitude, and a conditional probability amplitude for $\Phi^j(\dulr)$ and $\chi^j_{\dulr}(\dulq)$, leading to their identification as electronic and photonic wavefunctions. Here, it is important to 
note that unlike the electron that is a subatomic particle, the photon is an excitation of the quantized electromagnetic radiation in the cavity. Therefore, the concept of photonic wavefunction used here is meant to describe 
the electric displacement amplitude of the radiation field in the cavity (see section~(\ref{section:QED_Ham})). 

It can be proved~\cite{GG_PTRSA2014} that the photonic conditional wavefunction, $\chi^j_{\dulr}(\dulq)$, satisfies
\ben
 \label{eq:exact_photon}       
 \hat{H}^{ph,j}_{\dulr}\chi^j_{\dulr}(\dulq)\\= V^j_{e-ph}(\dulr) \chi^j_{\dulr}(\dulq),
\een
with the photonic Hamiltonian 
\ben
 \label{eq:photon_ham}
 \begin{split}
  \hat{H}^{ph,j}_{\dulr} =  \hat{H}_{\text{EM}} &+ \sum_{k=1}^{N_e}\frac{1}{m} \Big[\frac{(-i\nabla_k-{\bf S}^j_k(\dulr))^2}{2} \\
  &+ \Big(\frac{-i\nabla_k \Phi^{j}}{\Phi^{j}}+{\bf S}^j_k(\dulr)\Big)\left(-i\nabla_k-{\bf S}^j_k(\dulr)\right)\Big],
 \end{split}
\een
while the electronic wavefunction $\Phi^j(\dulr)$ satisfies a Schr\"odinger-like equation:
\bea
 \label{eq:exact_e} 
  \Bigl(\sum_{k=1}^{N_e}\frac{1}{2m}(-i\nabla_k+{\bf S}^j_k(\dulr))^2 +\hat{V}(\dulr)+\hat{W}_{ee}(\dulr) &+& \nonumber \\
   V^j_{e-ph}(\dulr)\Bigr)\Phi^j(\dulr)&=&E_j \Phi^j(\dulr),\nonumber \\            
\eea
where $m$ is the electronic mass. As it can be seen in Eq.~(\ref{eq:exact_e}), as a result of the coupling to the cavity photons, the electronic subsystem contains two additional potentials compared to the independent uncoupled electronic SE~(\ref{eq:ese}). 
The influence of electron-photon correlation on the $j$-th electronic state is formally exactly taken care of by addition of a scalar potential,   
\ben
 \label{eq:exact_epes}
 V^j_{e-ph}(\dulr) = \left\langle\chi^j_{\dulr} \right\vert\hat{H}^{ph,j}_{\dulr} \left\vert\chi^j_{\dulr}\right\rangle_\dulq, 
\een
and a vector potential,
\ben
 \label{eq:exact_evect}
 {\bf S}^j_k(\dulr)=\left\langle\chi^j_{\dulr}\right\vert\left.-i\nabla_k\chi^j_\dulr \right\rangle_\dulq, 
\een
to the uncoupled electronic SE~(\ref{eq:ese}). Here, $\langle ...|...|...\rangle_\dulq$ denotes an inner product over all photonic variables only.

Similar to the other extensions of the EF framework, the marginal electronic wavefunction and the conditional photonic wavefunction and their corresponding equations have the 
following properties:
\begin{itemize}
\item {Eqs.~(\ref{eq:exact_photon})-~(\ref{eq:exact_e}) are form-invariant under
the following gauge-like transformation,
\ben
 \label{eq:gt_wfs}
 \begin{split}
    &\chi^j_{\dulr}(\dulq)\rightarrow\tilde{\chi}^j_{\dulr}(\dulq)=\exp(i\theta_j(\dulr))\chi^j_{\dulr}(\dulq ) \nonumber \\ 
    &\Phi^j(\dulr )\rightarrow\tilde{\Phi}^j(\dulr )=\exp(-i\theta_j(\dulr ))\Phi^j(\dulr).
    \end{split}
\een
}
\item{The scalar potential given in Eq.~(\ref{eq:exact_epes}) is also gauge invariant under the above-mentioned gauge transformation while  
the vector potential is transformed as
\ben
{\bf S}^j_k(\dulr)\rightarrow\tilde{\bf S}^j_k(\dulr)={\bf S}^j_k(\dulr)+\nabla_k\theta_j(\dulr).
\een
}
\item{The wavefunctions $\chi^j_{\dulr}(\dulq)$ and $\Phi^j(\dulr)$ are unique up to this $(\dulr)$-dependent gauge transformation and yield the given solution, $\Psi^j(\dulr,\dulq )$,
of Eq.~(\ref{eq:se}) .}

\item{The electronic wavefunction, $|\Phi^j(\dulr )|^{2}=\int |\Psi^j(\dulr,\dulq )|^{2}d\dulq$, gives the probability density of finding
the electronic configuration $\dulr$ of the $j$-th correlated electron-photon state and the photonic conditional wavefunction,\\ $|\chi^j_{\dulr}(\dulq )|^{2}=|\Psi^j(\dulr,\dulq )|^{2}/|\Phi^j(\dulr )|^{2}$, 
provides the conditional probability of finding the displacement amplitudes of the cavity at $\dulq$ for a given electronic configuration $\dulr$. Furthermore, the exact electronic $N_e$-body current-density can be obtained 
from $\Im (\Phi^{j*}\nabla_k\Phi^j)+|\Phi^j(\dulr)|^{2}{\bf S}^j_k$. Therefore, $\chi^j_{\dulr}(\dulq)$ and $\Phi^j(\dulr)$ can be interpreted as photonic and electronic wavefunctions.}
\end{itemize}

One of the main results of this work is Eq.~(\ref{eq:exact_e}) that can be regarded as the {\it exact electronic equation} for the $j$-th electronic state of the correlated electron-photon system:
The Hamiltonian that is formed by adding the scalar potential (Eq.~(\ref{eq:exact_epes})) and the vector potential (Eq.~(\ref{eq:exact_evect})) (which are {\it unique} up to 
within a gauge transformation), to the uncoupled electronic Hamiltonian provides us with the $j$-th electronic state, $\Phi^j(\dulr)$, that yields the true electron ($N_e$-body) density and current density 
of the full electron-photon problem. 

The scalar electron-photon ({\it e-ph}) correlation potential~(\ref{eq:exact_epes}) can also be written as
\bea
 \label{eq:exact_epes2}
 V^j_{e-ph}(\dulr) &=& \left\langle\chi^j_{\dulr} \right\vert\hat{H}_\text{EM}(\dulq,\dulr) \left\vert\chi^j_{\dulr}\right\rangle_\dulq\nonumber\\
                          &+&\frac{1}{2m}\sum_{k=1}^{N_e}\left(\left\langle\nabla_k\chi^j_{\dulr}\vert\nabla_k\chi^j_{\dulr}\right\rangle_\dulq-{\bf S}^j_k(\dulr)^2\right), 
\eea
which for our purposes here has a more convenient form.
\begin{figure}
\begin{center}
\resizebox{0.45\textwidth}{!}{\includegraphics{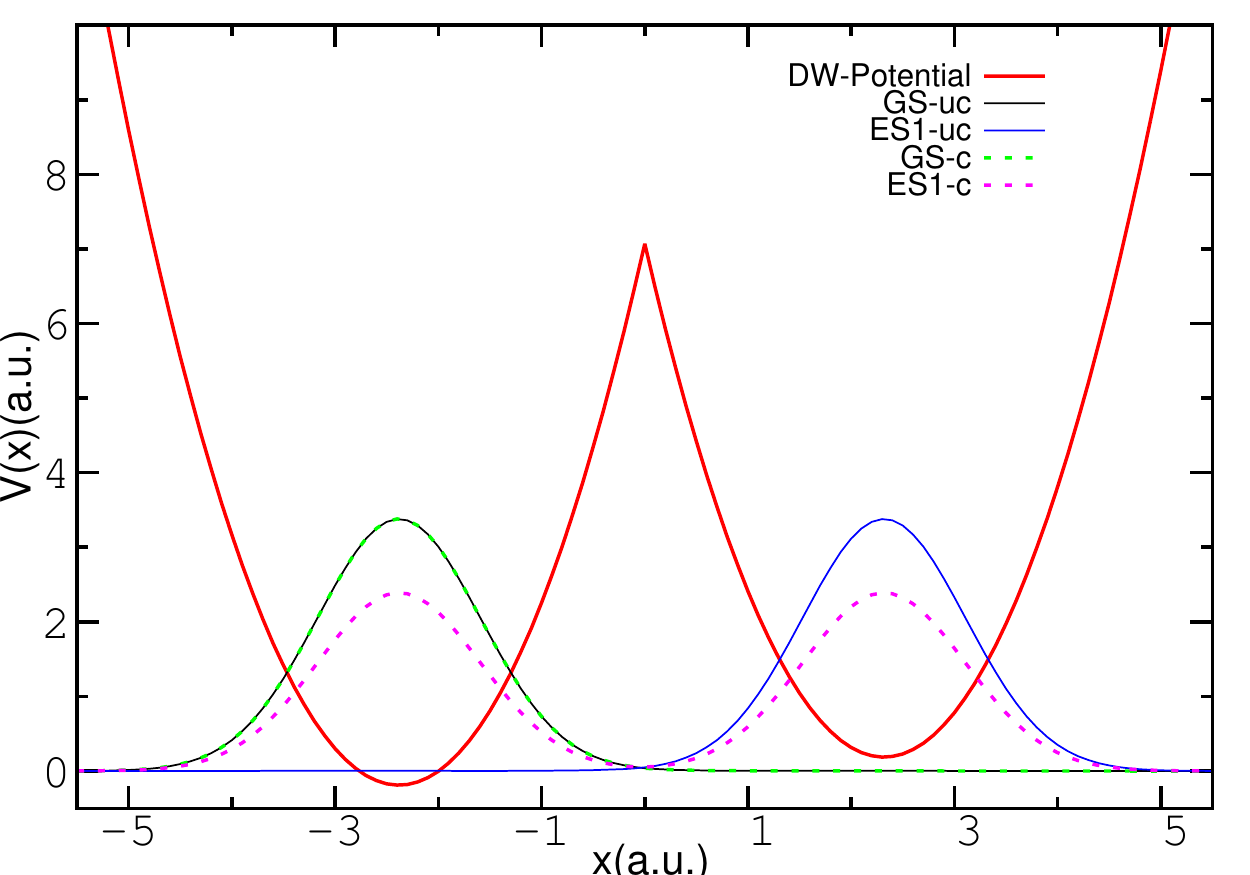}}
\end{center}
\caption{Asymmetric double-well potential (red) together with its ground-state density (black solid-line) and its 1st-excited state density (blue solid-line)  as well as the electronic ground-state (green dashed-line) 
and 1st-excited state (magenta dashed-line) densities of the complete electron-photon system. The densities have been enlarged four times. The acronyms "c" and "uc" on the plot-label stand for "coupled" and "uncoupled" 
respectively.}
\label{fig:fig1}
\end{figure}
\section{Example: photon-assisted delocalization of the electronic states}
In this section, we investigate a particular situation in which the character of the electronic excited states of the system undergoes a fundamental change through electron-photon coupling, 
i. e. they become delocalized as a result of the coupling to a cavity mode with a resonance frequency. Our study is based on a model system in which 
we consider a single electron in an Asymmetric Double-Well (ADW) potential 
\ben
\label{eq:ADWpot}
V(x)=\frac{1}{2} \omega_e^2 (|x|-a)^2 + E x,
\een
with the Hamiltonian
\ben
\label{eq:DWeH}
\hat{H}^\text{ADW}_{e} = -\frac{1}{2}\frac{\partial^2}{\partial x^2} + V(x)
\een
Here, $\omega_e=1.6$ (a.u.) , $a=2.35$ (a.u.) , the static electric field $E=0.08$ (a.u.), and the electronic mass $m$ is set to $1$. These parameters are chosen such that this electronic system is 
practically a two-level system. In Fig.~\ref{fig:fig1} the asymmetric double-well potential (Eq.~\ref{eq:ADWpot}) and the first two electronic states are shown. 
Due to the asymmetric nature of the potential the electronic ground state and the first excited state are localized in one of the wells that are centered at $\pm a$ with a relatively small overlap, hence, the probability of inter-well tunneling is very small. When this system is coupled to a single-photon mode of a cavity with the frequency $\omega_{c}$ and coupling constant of $\lambda_{c}$, the full description 
is given by the electron-photon Hamiltonian 
\ben
\label{eq:Ham_DWpC}
 \hat{H}_\text{tot} = \hat{H}^\text{ADW}_{e} + \hat{H}_\text{EM},
\een
that now contains
\ben
\label{eq:Hq}
\hat{H}_\text{EM} = \frac{-1}{2}\frac{\partial^2}{\partial q^2}+\frac{1}{2}\omega_{c}^2\left(q-\frac{\lambda_{c}}{\omega_{c}}x\right)^{2},
\een
as a result of the quantization of the electromagnetic field and the subsequent electron-photon coupling (\ref{Hem-PZW}). If the frequency of the cavity-photons 
is tuned to bring the first two electronic states of the asymmetric double-well into resonance, the primarily localized electronic excited state becomes delocalized (see Fig.~\ref{fig:fig1}) independent of the value of the coupling constant. The resonance frequency depends on $\lambda_c$: for small $\lambda_c$-s, $\omega_c\sim2 E a$ and for larger coupling constants, the $\lambda_c$-dependence becomes more pronounced.  In Table.~(\ref{table:T1}) the resonance frequencies corresponding to three values of $\lambda_c$ that are considered in this work are given. 
\begin{table}[]
\centering
\label{table:T1}
\begin{tabular}{|c|c|c|c|}
\hline
 $\lambda_c$ & 0.1 & 0.5 & 0.9 \\
 \hline 
 $\omega_c$ & 0.37676260 & 0.39495042 & 0.43442993  \\
 \hline 
\end{tabular}
\caption{The coupling constants and the corresponding resonance frequencies implemented in this work in atomic units .} 
\end{table}

In the following, we investigate how this photon-assisted delocalization of the electronic excited state can be captured by adding the {\it e-ph} correlation potential~(\ref{eq:exact_epes2}) to the electronic 
Hamiltonian~(\ref{eq:DWeH}). Here, we are not aiming at solving the EF equations~(\ref{eq:exact_photon})-~(\ref{eq:exact_e}). In fact, these coupled equations need to be solved self-consistently that seems to be somehow more 
complicated than solving the full SE~(\ref{eq:se}). Therefore, we solve the full SE~(\ref{eq:se}) and similar to our previous studies~\cite{AMG_PRL2010,AMG_JCP2013,AASG_PRL2013} extract the numerically exact 
{\it e-ph} potential by inverting the full SE. In addition, we derive an approximate analytical expression for the exact {\it e-ph} correlation potential~(\ref{eq:exact_epes}) and show how step and peak features of the 
{\it e-ph} correlation potential are captured analytically. These features lead to the polaritonic localization of the electronic ground-state and delocalization of the electronic excited-states.  

For the one dimensional model (Eq.~\ref{eq:DWeH}), the {\it e-ph} correlation potential reads   
\ben
 \label{eq:exact_epes2}
 V^j_{e-ph}(x) = \left\langle\chi^j_{x} \right\vert\hat{H}_\text{EM}\left\vert\chi^j_{x}\right\rangle_q+\frac{1}{2m}\left\langle\frac{\partial\chi^j_{x}}{\partial x}\vert\frac{\partial\chi^j_{x}}{\partial x}\right\rangle_q,
\een
while the vector potential can be gauged away~\cite{AMG_JCP2012}.
\subsection{Combining atomic orbitals and displaced harmonic oscillator basis}
\begin{figure}
\begin{center}
\includegraphics[width=0.45\textwidth]{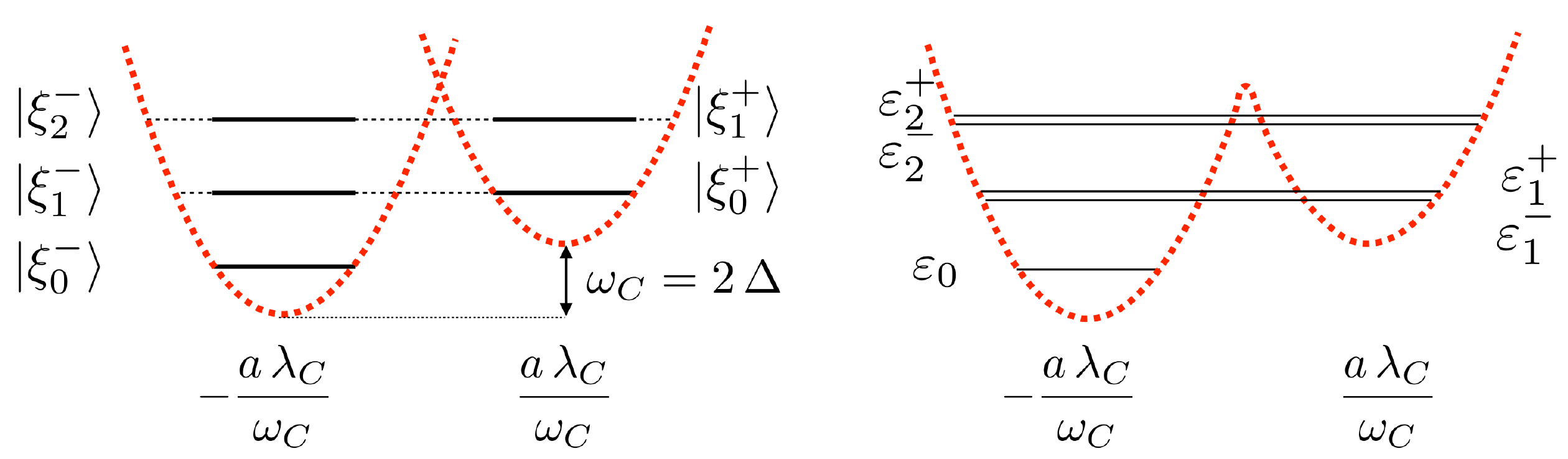}
\end{center}
\caption{(Left) Schematic representation of the displaced harmonic oscillator basis with the usual eigenstates. Eigenstates with the energy quantum number $N$ on the left are degenerate with the Eigenstates with energy 
quantum number $N-1$ on the right. (Right) Schematic representation of the (diabatic) approximation in which the $N$-th state of the displaced oscillator on the left is allowed to mix with the $(N-1)$-th state of the displaced 
oscillator on the right (similar to the symmetric case of Irish et al.~\cite{irish2005dynamics}).}
\label{fig:fig2}
\end{figure}
In order to work out an accurate analytical approximation for the complete electron-photon wavefunction, $\Psi^j(x,q)$, that is the $j$-th eigenstate of Eq.~(\ref{eq:Ham_DWpC}), we first expand it in a basis set. Given the fact that 
the uncoupled electronic system is practically a two-level system with two localized states, it can be accurately described by localized atomic orbitals (AO) that are the ground states of the harmonic potentials 
centered at $\pm a$, i. e.,
\ben
\label{eq:AO_se}
\left[\hat{T}_e+\hat{V}^{\pm}_{at}\right]\phi^{\pm}(x) = E^{\pm} \phi^{\pm}(x)
\een
where $\hat{V}^{\pm}_{at}=\frac{1}{2}\omega_e^2\left(x \mp a\right)^2$. To form our basis set we combine these atomic orbitals with the so called displaced harmonic oscillator (DHO) basis. The DHO basis  
are the eigenstates of two harmonic oscillator Hamiltonians that are centered at $\pm \frac{a\lambda_c}{\omega_c}$, i. e.,  
\begin{equation}
  \hat{H}_q^{\pm} \xi^{\pm}_N(q) = V_{\textrm{ad}}^{N \pm } \xi^{\pm}_N (q) 
\end{equation}
where $\hat{H}_q^{\pm} = \frac{-1}{2}\frac{\partial^2}{\partial q^2}+\frac{1}{2} \omega_{c}^{2}\left( q\, \mp \frac{\lambda_c}{\omega_c} a\right) ^2 \pm\Delta$ and 
$V_{\textrm{ad}}^{N \pm } = \pm \Delta + (N + \frac{1}{2}) \omega_c$~(with $\Delta=E\, a$). We then choose the localized basis as products of the AO on the left (right) and the DHO states on the left (right)
, $\{\xi^{-}_N (q) \phi^{-}(x)\}$($\{\xi^{+}_N (q)  \phi^{+}(x)\}$) and expand the full wavefunction in terms of them, i. e., 
\begin{equation}
\label{eq:expansion}
\Psi(x,q) = \sum_{N=0}^{\infty} \left[ A _{N}^{-} \xi^{-}_N (q)  \phi^{-}(x) + A _{N}^{+} \xi^{+}_N (q)  \phi^{+}(x) \right], 
\end{equation}
\begin{figure}
\begin{center}
\includegraphics[width=0.5\textwidth]{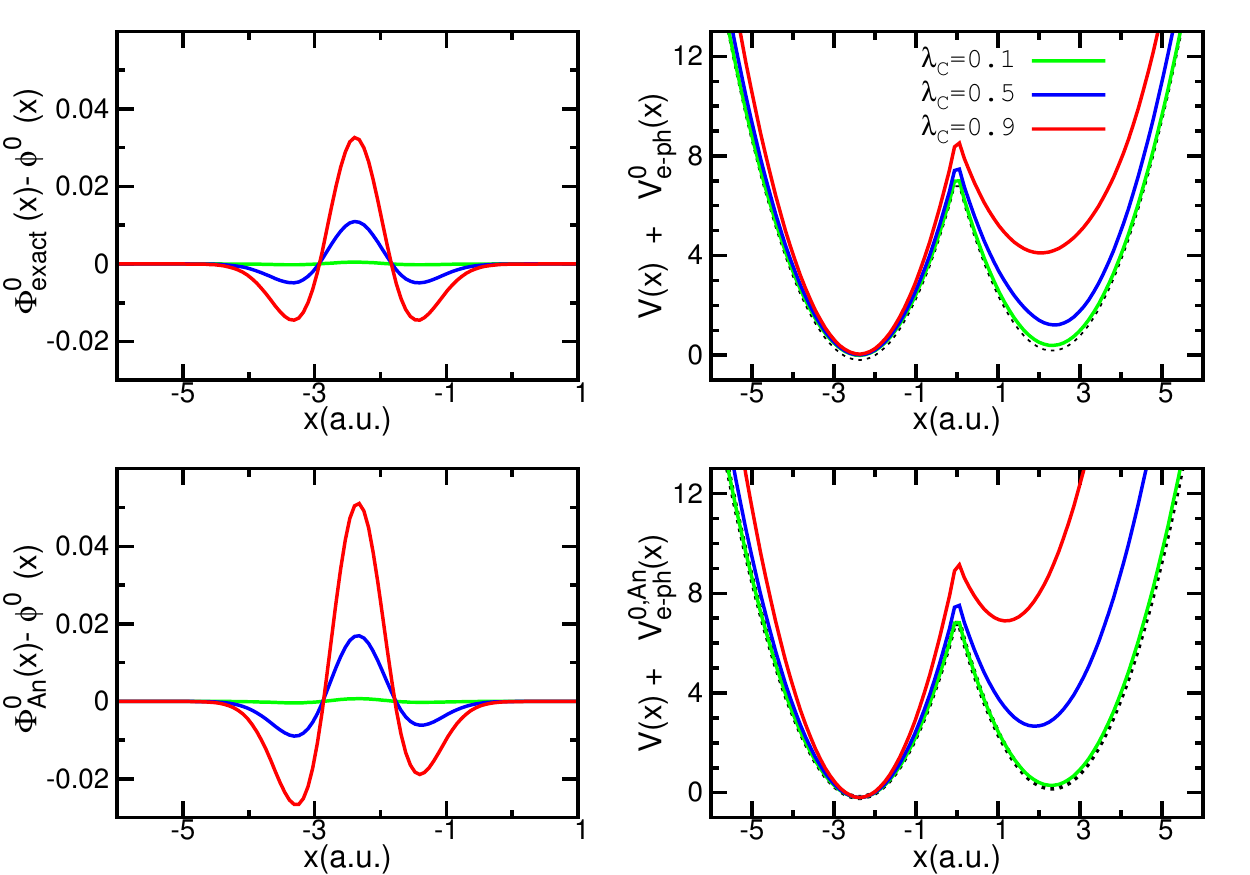}
\end{center}
\caption{Right: Ground-state full electronic potential including the {\it e-ph} correlation potential for various $\lambda_c$-s as indicated on the plots calculated from numerically exact solution of SE (top) and 
from our analytical approximation (bottom). The asymmetric double-well potential of the uncoupled electronic system has been plotted (black dashed-line) in both for a reference. Left: the difference between the  
electronic density of the uncoupled electronic system with the electronic density of coupled electron-photon system calculated from numerically exact solution of SE (top) and from our Analytical Approximation (AN) (bottom) 
for various $\lambda_c$-s as indicated on the plots.}  
\label{fig:fig3}
\end{figure}
where $A _{N}^{-}$($A _{N}^{+}$) are the expansion coefficients of the left (right) states. This is analogous to the expansion of the full molecular wavefunction in terms of a diabatic basis.
Our choice of basis was inspired by the work of Irish et al.~\cite{irish2005dynamics} in which they implemented the DHO to solve the two-level Rabi model. 
We now plug the expansion~(\ref{eq:expansion}) into the complete electron-photon SE~(\ref{eq:se}) with the Hamiltonian~(\ref{eq:Ham_DWpC}) and project onto $ {\langle \xi^{-}_M}\phi^{-}\vert$ (and  ${\langle \xi^{+}_N}\phi^{+}\vert$)) 
that leads to
 \bea
 \label{eq:projection}
 & &A_{M}^{-} \left[ \alpha + \hat{V}_{\textrm{add}}^{M-}  \right]+\nonumber\\
 & &\sum_{N=0}^{\infty} A _{N}^{+} \left[ \beta \langle\xi^{-}_M(q)  \vert \xi^{+}_N(q)\rangle_{q}  + \hat{V}_{\textrm{add}}^{N+} \, \langle\xi^{-}_M(q)  \vert \xi^{+}_N(q)\rangle_{q} S \right] \nonumber\\
 & & \,\,\,\,\,\,\,\,\,\,\,\,\,\,\,\,\,\,\,\,\,\,\,\,\,\,\,\,\,\,\,\,\,\,\,\,\,\,\,\,\,\,\,= E \left [A_M^{-} +  \sum_{N=0}^{\infty} A _N^{+} \, \langle\xi^{-}_M(q)  \vert \xi^{+}_N(q)\rangle_{q} S \right],\nonumber\\
 \eea
where 
\bea
\alpha &=&  \langle \phi^{\pm}(x)\vert  \left[ \hat{T}_e + \frac{1}{2} \omega_e^2 (|x|-a)^2 \right]  \vert \phi^{\pm}(x)\rangle, \nonumber\\ 
\beta &=& \langle \phi^{\pm}\vert \left[ \hat{T}_e + \frac{1}{2} \omega_e^2 (|x|-a)^2 \right] \vert  \phi^{\mp} \rangle,\nonumber \\ 
S &=& \langle \phi^{-}\vert  \phi^{+} \rangle,
\eea
are the typical elements of the LCAO technique. The outcome of the projection onto ${\langle \xi^{+}_N}\phi^{+}\vert$ is symmetric with Eq.~(\ref{eq:projection}).   
In Fig.~\ref{fig:fig2} we present the DHO basis schematically. As it can be seen in the figure, for the resonance frequencies the states of the left DHO with the energy quantum number $N^{-}$ are 
degenerate with the states of the right DHO with the energy quantum number $(N-1)^{+}$. In this work we only allow for the mixing of these states and neglect the coupling to the other states 
that can be regarded as an adiabatic approximation. We discuss how to go beyond this limit in a different publication~\cite{AKT_IP2018}.

\subsubsection{Ground state}
\begin{figure}
\begin{center}
\includegraphics[width=0.5\textwidth]{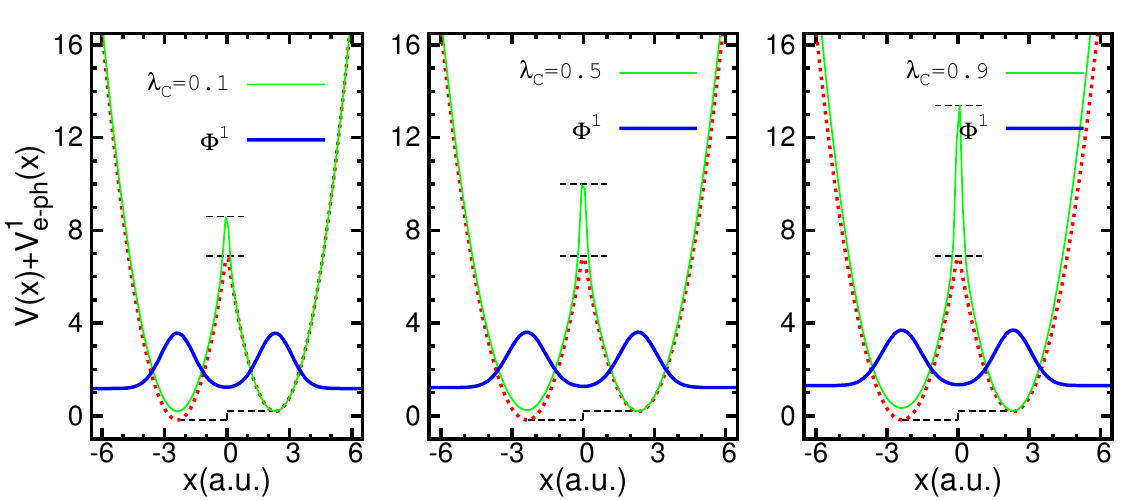}
\end{center}
\caption{First excited-state electronic densities (blue solid-line) for three different $\lambda_c$-s (indicated on the plots) together with their corresponding full electronic potentials (green solid-line), obtained from 
the numerical integration of the electron-photon SE~(\ref{eq:se}). The electronic potential of the uncoupled electronic system (asymmetric double-well) is plotted (red dashed-line) as a reference. Step and peak features 
are highlighted on all the plots.}
\label{fig:fig4}
\end{figure}
The full electronic potential of the EF framework for the ground-state, $V(x)+V^{0}_{e-ph}(x)$, with the {\it e-ph} correlation potential~(\ref{eq:exact_epes2}) obtained from the numerically exact solution of SE 
of combined systems~(\ref{eq:se}) has been plotted in Fig.~\ref{fig:fig3} (top-right) for $\lambda_c = 0.1\, ,0.5\, ,0.9$. As it can be seen in this plot, for smallest coupling constant $\lambda_c=0.1$ the 
full electronic potential only slightly differs from the asymmetric double-well potential of the uncoupled electronic system. By increasing $\lambda_c$, the well on the right side is lifted up and both wells are squeezed 
leading to a polaritonic squeezing of the electronic states compared to the electronic states of the uncoupled electronic system as it is clear in the top-left panel of the Fig.~\ref{fig:fig3}, hence, the larger the coupling 
constant, the more squeezed the electronic density. 

We now turn to our approximate evaluation of the correlated {\it e-ph} potential to see whether this effect is reflected in our approximation and if yes how? 
Due to the asymmetry of the potential in Hamiltonian~(\ref{eq:DWeH}), there is no state from the right side ($+$) in resonance with the localized state $\Phi^{-}\xi^{-}_{0} (q)$ at the left side. Therefore, the conditional 
photonic wavefunction of the ground state can be approximated as $\xi^{-}_{0} (q)$. The corresponding {\it e-ph} correlation potential reads
\ben
 \label{eq:epes_gs}
 V^{0}_{e-ph}(x) = \left\langle\xi^{-}_{0} (q) \right\vert\hat{H}_\text{EM}\left\vert\xi^{-}_{0} (q)\right\rangle_q = \frac{\omega_c}{2} + \frac{\lambda_c^2}{2}\left(x+a\right)^2,
\een
as the second term in Eq.~(\ref{eq:exact_epes2}) is zero. The final outcome of Eq.~(\ref{eq:epes_gs}) is obtained by replacing $\hat{H}_\text{EM}$ with\\ 
\ben
\hat{H}_q^{-}+\frac{1}{2}\, \omega_{c}^{2}\left[\left(q-\frac{\lambda_{c}}{\omega_{c}}x\right)^{2}-\left( q\, + \frac{\lambda_c}{\omega_c} a\right) ^2\right]+\Delta. \nonumber
\een
In Fig.~\ref{fig:fig3} (bottom-right), we present the full electronic potential that is resulted from addition of this analytically approximate {\it e-ph} correlation potential~(\ref{eq:epes_gs}) to the asymmetric double-well potential 
of the uncoupled electronic system. Furthermore, we implement these potential to calculate the electronic states from Eq.~(\ref{eq:exact_e}) and show the difference between the  
electronic densities of the uncoupled electronic system with the approximate electronic densities of coupled electron-photon system in Fig.~\ref{fig:fig3} (bottom-left). It can be seen that the potentials and densities 
corresponding to the approximate analytical {\it e-ph} correlation potentials follow the same trend as the numerically exact results and exhibit the main features, i.e., the right well is elevated while both wells are 
squeezed. The densities are also squeezed compared to the electronic density of the uncoupled electronic system, following the same trend as the exact electronic densities. Although, in both cases of approximate potentials and electronic 
densities the squeezing is more exaggerated. The approximate {\it e-ph} correlation potential also shifts up the right well fairly above the exact results. However, given the fact that the most 
simple approximation for the conditional photon wavefunction has been implemented here, the agreement with the exact result is fulfilling. 
\begin{figure}
\begin{center}
\includegraphics[width=0.5\textwidth]{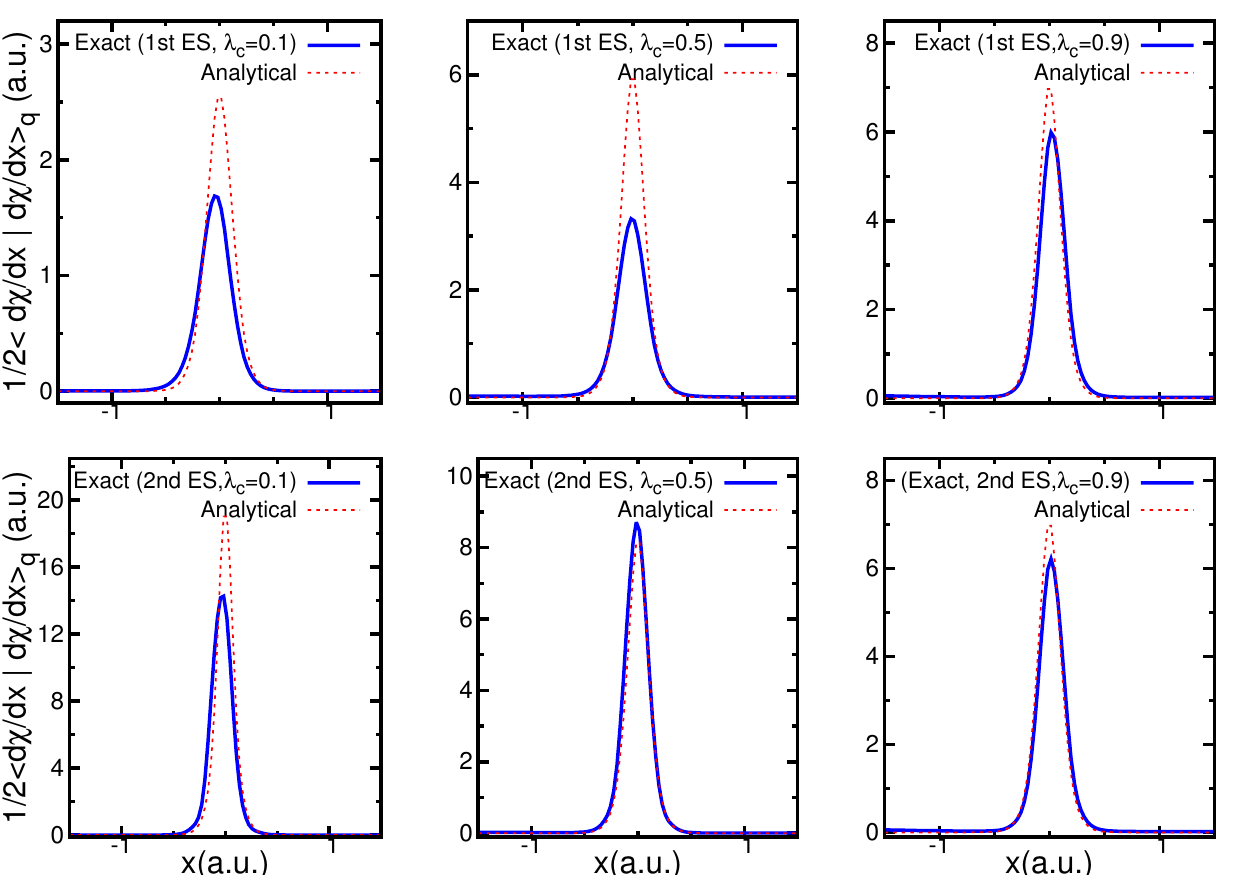}
\end{center}
\caption{Second term of Eq.~(\ref{eq:exact_epes2}) for the first (upper panel) and second (lower panel) excited state calculated from the numerically exact solution of SE (blue solid-line) and from the 
approximate analytical expression given in Eq.~(\ref{eq:nabla_nabla}) (red dashed-line).}
\label{fig:fig5}
\end{figure}

\subsubsection{Excited states}
Now we turn to investigate the {\it e-ph} correlation potential of the electronic excited-states of the coupled electron-photon system for the resonance photon frequencies. Contrary to the ground-states, the excited-states electronic 
densities are totally delocalized. In Fig.~\ref{fig:fig4} we have plotted the first excited-state electronic densities for three different coupling constants together with their corresponding full electronic potentials, 
$V(x)+V^{1}_{e-ph}(x)$ obtained from the numerical integration of the electron-photon SE~(\ref{eq:se}). As it is also highlighted in the figure, the addition of the {\it e-ph} correlation potential~(\ref{eq:exact_epes2}) to the electronic potential $V(x)$, significantly modifies the electronic potentials 
in two ways: First, it brings the two-wells to the same level, symmetrizing the asymmetric electronic potential which is highlighted on the plots with steps. Second, it increases the barrier between the two wells that
stabilizes the delocalization of the electronic densities on the both sides of the double-well potential. In addition, the wells are squeezed as the coupling constant increases that leads to polaritonic squeezing of the delocalized 
electronic excited states. 
\begin{figure}
\begin{center}
\includegraphics[width=0.5\textwidth]{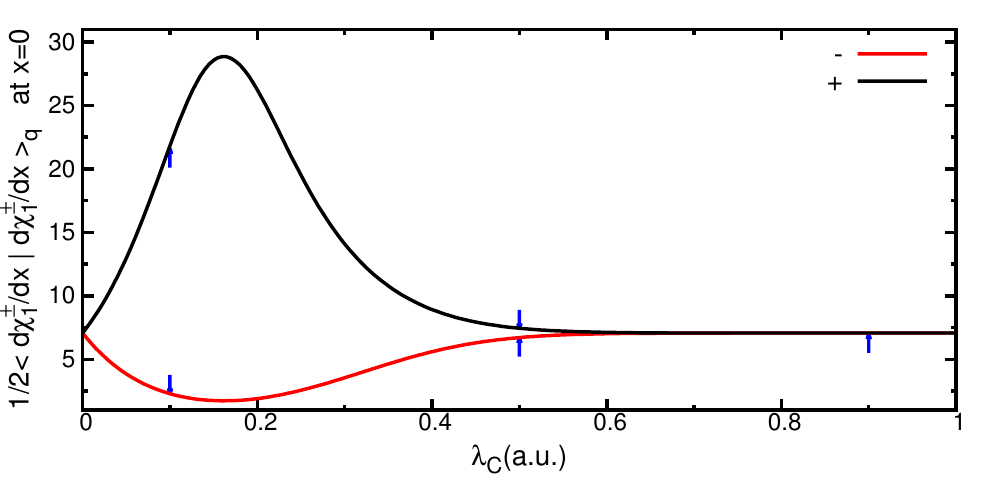}
\end{center}
\caption{The height of the peak at $x=0$ as a function of $\lambda_c$ estimated by Eq.~\ref{eq:nabla_nabla_0}. The blue arrows indicate the height-estimations for the coupling constants $\lambda_c$ used in this work.}
\label{fig:fig6}
\end{figure}

We shall now investigate how these features are captured by our approximate analytical treatment. Here we note that while we give general analytical expressions for the {\it e-ph} correlation potential, our discussions are only 
focused on $N=1$. Furthermore, our results are valid as long as the energies of coupled electron-photon system is well bellow the second excited state of the uncoupled electronic system.

As we only allow for the mixing of the states with the same energy in the expansion~(\ref{eq:expansion}), within our approximate treatment, the full electron-photon excited-states that are the eigenstates of Eq.~(\ref{eq:Ham_DWpC}) 
may be written as  
\begin{equation}
\label{eq:el_ph_wf}
\Psi^{\pm}_{N}(x,q)= \frac{1}{\sqrt{2\nu^{\pm}_N}} \left[ \xi^{-}_{N}  \phi^{-} \pm  \xi^{+}_{N-1}  \phi^{+} \right],
\end{equation}
with the corresponding excited-state energies
\begin{equation}
E^{\pm}_N= N \, \omega_c+ \frac{\alpha \pm \beta \langle \xi_{N}^{-}  \vert\xi_{N-1}^{+}  \rangle_q}{\nu^{\pm}},
\end{equation}
where
\begin{equation}
 \nu^{\pm} = 1\pm S  \, \langle \xi_{N}^{-}  \vert\xi_{N-1}^{+}  \rangle_q, 
\end{equation}
and
\ben
\langle \xi_{N}^{-}  \vert\xi_{N-1}^{+}  \rangle_q = \exp(- \frac{a^2 \, \lambda_c^2}{2 \, \omega_c^2}) \left( \frac{- a \,  \lambda_c} {\omega_c} \right)  \sqrt{\frac{1}{N}}\,\mathcal{L}_{N}^{1}\left(\frac{a^2 \lambda_{c}^2}{ \omega_c^2}\right).
\een
Here $\mathcal{L}_{i}^{j}$ is an associated Laguerre polynomial~\cite{irish2005dynamics}. From this electron-photon wavefunctions we can obtain the corresponding electronic wavefunction,
\ben
 \Phi^{\pm}_{N}(x) =  \frac{1}{\sqrt{2\nu^{\pm}_N}} \left[ \vert \phi^{-} \vert^2 + \vert \phi^{+} \vert^2  \pm 2 \langle \xi^{-}_{N}  \vert\xi^{+}_{N-1} \rangle_{q} \phi^{-} \phi^{+}  \right]^{1/2},
 \label{eq:phi_app}
\een
and conditional photonic wavefunction,
\ben
 \chi^{\pm}_{N} (q\vert x) = \frac{1}{\sqrt{2}}  \frac{\xi^{-}_{N}  \phi^{-} \pm  \xi^{+}_{N-1}  \phi^{+} }{\Phi^{\pm}_{N}(x)}.     
 \label{eq:chi_app}
\een
Then we simply implement the conditional photonic wavefunction to derive the {\it e-ph} correlation potential~(\ref{eq:exact_epes2}). Here, we first discuss the expression of the second term in~(\ref{eq:exact_epes2}) that has 
the following analytical form in our approximation (considering $m=1$):
\ben
\label{eq:nabla_nabla}
\frac{1}{2}\langle  \partial_x \chi^{\pm}_{N} \vert  \partial_x \chi^{\pm}_{N} \rangle_q = \frac{a^2\,\omega_e^2\, \vert \phi^{+} \vert^2 \vert \phi^{-} \vert^2  (1-\langle\xi^{-}_{N} \vert\xi^{+}_{N-1} \rangle_{q}^2) }{2\vert\Phi_{N}^{\pm}\vert^4}. 
\een
\begin{figure}
\begin{center}
\includegraphics[width=0.5\textwidth]{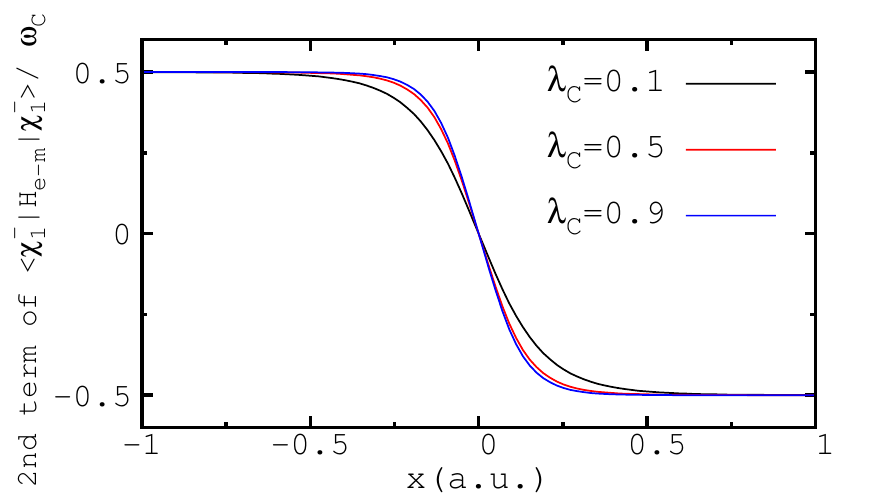}
\end{center}
\caption{Estimation of the step from the second term of Eq.~\ref{eq:e_approx} for three different $\lambda_c$-s}
\label{fig:fig7}
\end{figure}

In Fig.~\ref{fig:fig5}, we plot the expression above for the first (upper panel) and second (lower panel) excited state together with numerically exact results for the same term (second term of~(\ref{eq:exact_epes2})). As it appears from the figure, our analytical expression gives a peak in the same position as the numerically exact results, reproducing the peak feature of the
{\it e-ph} correlation qualitatively well. Here we can further simplify the expression of the peak~(\ref{eq:nabla_nabla}) for the center of the peak that is located at the crossover of the two AOs that happens to be at $x=0$. 
Hence, the approximate expression predicts the height of the peak at $x=0$ as  
\ben
\label{eq:nabla_nabla_0}
\frac{1}{2} \langle  \partial_x \chi^{\pm}_{N} \vert  \partial_x \chi^{\pm}_{N} \rangle_q\vert_{x=0} = \frac{a^2\,\omega_e^2}{2}\left(\frac{1\mp\langle\xi^{-}_{N} \vert\xi^{+}_{N-1} \rangle_{q}}{1\pm\langle\xi^{-}_{N} \vert\xi^{+}_{N-1} \rangle_q}\right). 
\een
In Fig.~\ref{fig:fig5} we have plotted this for $N=1$ that leads to a prediction for the height of the peak for the first ($1^{-}$) and second ($1^{+}$) excited states. According to this expression, 
the height of the peaks for the first two excited states progress in different directions. While the height of the peak for the the lower state ($-$) descends to a minimum (around $\lambda_c=0.17$) the height of the peak for 
the upper-state ($+$) ascends to a maximum around the same point. However, for large $\lambda_c$-s the heights of both states converge to $\frac{a^2\,\omega_e^2}{2}$.
\begin{figure}
\begin{center}
\includegraphics[width=0.5\textwidth]{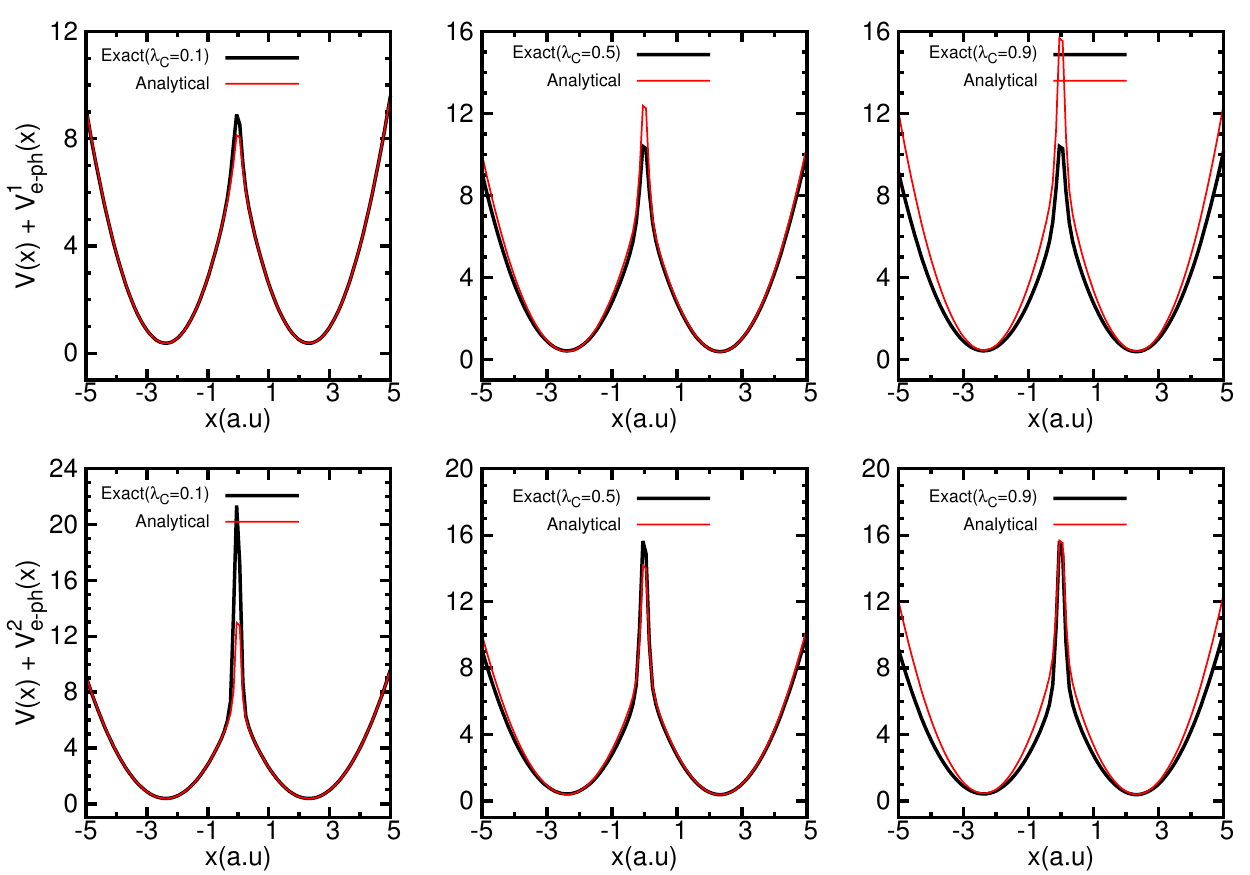}
\end{center}
\caption{Full electronic potentials obtained from an approximate analytical expression (red dashed-line) versus the ones calculated numerically exactly (blue solid-line) for three different $\lambda_c$-s as indicated in the plots}
\label{fig:fig8}
\end{figure}

Now let us investigate the analytical expression for the first term of Eq.~(\ref{eq:exact_epes2}) that within our approximation is
\begin{equation}
\label{eq:e_approx}
\left\langle\chi^{N\pm}_{x} \right\vert\hat{H}_\text{EM}\left\vert\chi^{N\pm}_{x}\right\rangle_q =   N \omega_{c} + \frac{\omega_{c}}{4}\left(\frac{\vert \phi^{-} \vert^2 - \vert \phi^{+} \vert^2 }{\vert\Phi_{N}^{\pm}\vert^2} \right)  + f(\lambda_{c})
\end{equation}
where
\bea\label{eq:e_approx_ldependent}
& &f(\lambda_{c}) = \frac{\lambda_{c}^2 a \, x}{2} \left(\frac{\vert \phi^{-} \vert^2 - \vert \phi^{+} \vert^2 }{\vert\Phi_{N}^{\pm}\vert^2} \right) + \nonumber\\
& &\frac{\lambda_{c}^2 x^2}{2}+ \frac{\lambda_{c}^2 a^2}{2} \left(\frac{\Phi_{N}^{\mp}}{\Phi_{N}^{\pm}}\right)^2 \mp  \frac{\lambda_{c} \, \omega_{c} \, x   \langle \xi^{-}_{N} \vert q \vert \xi^{+}_{N-1}   \rangle_{q}}{\vert\Phi_{N}^{\pm}\vert^2} \phi^{-} \phi^{+}\nonumber\\
\eea
While the first term of~(\ref{eq:e_approx}) gives a constant general shift, the second term gives a step-like function with the height equal to $\omega_c$ ($=2 \Delta$) as it is shown in Fig.~\ref{fig:fig7} and as it can be seen 
in the figure, the larger the coupling constant $\lambda_c$ the sharper the step around $x=0$. This proves how this approximation can capture yet another essential feature of the exact {\it e-ph} correlation potential that was shown schematically in 
Fig.~\ref{fig:fig4} and was discussed earlier in this section. 
Finally, as the concluding result of this work, we present the full electronic potentials $V(x)+V^{1(2)}_{e-ph}$ obtained analytically approximately versus the ones calculated numerically exactly in Fig.~\ref{fig:fig8} for three different $\lambda_c$-s 
as indicated in the figure. As it was discussed previously and can also be seen in Fig.~\ref{fig:fig8}, the approximate analytical expression for the {\it e-ph} correlation potential captures the essential step and peak
features of the exact {\it e-ph} correlation very well. It also captures the squeezing of the wells but overestimates this feature as $\lambda_c$ increases.     

\section{Summary and Outlook}
We have extended the EF framework to study the electronic states of the correlated electron-photon systems and have shown that within this new approach to correlated electron-photon states, the electronic states are 
uniquely determined by addition of a scalar potential ({\it e-ph} correlation potential) and a vector potential to the uncoupled electronic Hamiltonian. For a one dimensional asymmetric double-well potential coupled to a 
single photon mode of a cavity with resonance frequencies, we have calculated the exact {\it e-ph} correlation potential numerically exactly and discussed their significant features namely steps, intra-well peaks and 
squeezing of the wells of the  double-well potential. These features of the {\it e-ph} correlation potential are responsible for the polaritonic squeezing of the electronic ground-state and photon-assisted delocalization 
as well as polaritonic squeezing of the electronic excited-states. Although not directly related, the step-and-peak structure of the {\it e-ph} correlation potential investigated in this work is reminiscent of 
the step-and-peak structure of the Kohn-Sham potential of density functional theory in the dissociation limit~\cite{helbig2009exact,Gritsenko1997,tempel2009revisiting}. 

We have furthermore derived an approximate analytical expression for the {\it e-ph} correlation of the model system studied, by extending the atomic orbitals via combining them with displaced harmonic oscillator states. 
In the case of the ground electronic state of the coupled electron-photon system, we have shown how our analytical approximation captures the key features of the exact {\it e-ph} correlation potential, i. e., squeezing of 
both wells and the elevation of the right well that leads to the polaritonic squeezing of the electronic ground-states in the left well which is enhanced as the coupling constant increases. In the case of the first two 
excited states, the analytical approximation reproduces the step and peak features of the exact {\it e-ph} correlation potential that are essential to capture the photon-assisted delocalization of the electronic 
excited-states. In the case of the ground-state and the first two excited state discussed in this work, we have shown that while the analytical approximation captures the polaritonic squeezing of the electronic states that is 
enhanced as $\lambda_c$ increases, it overestimates this effect. In our upcoming work we will discuss how to go beyond the approximation presented here~\cite{AKT_IP2018}. One of the main motivation of these analytical 
investigations  is to set the stage to utilize the time-dependent EF framework for studying the correlated electron-photon dynamics~\cite{AKT2_IP2018}. This approach to coupled electron-photon dynamics is complementary 
to the recently developed TDDFT approach to cavity QED. Indeed, the {\it e-ph} correlation potential which was the heart of our investigation in this work is closely related to the correlation potential of the cavity QED (TD)DFT, therefore,
the features of the {\it e-ph} correlation potential we discussed in this work together with the analytical approximation of the potential presented here are particularly relevant for developing the 
cavity QED (TD)DFT exchange-correlation functionals~\cite{FSR18,DFRR17}. Another interesting avenue to explore is the connection between the approach proposed here and  
the Born-Huang expansion approach for the cavity QED that has been proposed very recently~\cite{SRR18}.

\section{Acknowledgments}
It is our great pleasure to dedicate this work to Hardy Gross who has revived and established the Exact Factorization approach through his persistent developments and exceptional presentations. We wish Hardy the best 
for the decades to come!

A. A.  and  E. K.  acknowledge funding from the European Uninos Horizon 2020 research and innovation programme under the Marie-Sklodowska-Curie grant agreement  no. 702406 and 704218, respectively.
I. T. acknowledges the funding from Spanish Ministerio de Economía y Competitividad (MINECO), project No. FIS2016-79464-P and by the “Grupos Consolidados UPV/EHU del Gobierno Vasco” (Grant No. IT578-13).
 
\addcontentsline{toc}{section}{References}
\bibliographystyle{ieeetr}
\bibliography{./factorization,./refs}

\end{document}